\documentstyle[preprint,prb,aps,epsf]{revtex}
\begin{document}
\draft
\title{
Elastic Properties of Carbon Nanotubes and Nanoropes
}
\author{Jian Ping Lu}
\address{
Department of Physics and Astronomy \\
University of North Carolina at Chapel Hill \\
Chapel Hill, North Carolina 27599 \\
jpl@physics.unc.edu
}
\date{\today}
\maketitle
\begin{abstract}
Elastic properties of carbon nanotubes and nanoropes
are investigated using an empirical force-constant model.
For single and multi-wall nanotubes
the elastic moduli are shown to be insensitive
to details of the structure such as the helicity, the
tube radius and the number of layers.
The tensile Young's modulus and the torsion
shear modulus calculated are comparable to that of the diamond,
while the the bulk modulus is smaller.
Nanoropes composed of single-wall nanotubes possess
the ideal elastic properties of high
tensile elastic modulus, flexible, and light weight.

\end{abstract}
\bigskip

\narrowtext
{\hspace{-0.5in}\large\bf I. Introduction}

The discoveries of carbon nanotubes\cite{iijima}
and the new efficient method of producing them\cite{thess}
stimulate a great interest in these novel materials.
The electronic\cite{hamada} and magnetic properties\cite{lu1}
of nanotubes depend sensitively on the structural details such as
the tube radius and the helicity.
It has been speculated that nanotubes also posses
novel mechanical properties.
Recent measurements have inferred
a Young's modulus that is several times 
that of the diamond.\cite{ebbesen}

The mechanical properties of small single-wall nanotubes
have been studied by several groups using molecular 
dynamics simulations.\cite{brenner,bernholc}
A Young's modulus several times greater than that of the
diamond was predicted.
However, those calculations were restricted to small single-wall
nanotubes of few \AA\ in radius.
Most samples of nanotubes are either multi-wall
or crystalline ropes of single-wall nanotubes.

A practical method of investigating elastic properties
is to use the empirical force-constant model.
The phonon spectrum and elastic properties of the graphite
has been successfully calculated using such 
models.\cite{jishi}
The similarity in local structure between the graphite
and the nanotubes ensure that 
a similar model is applicable for nanotubes.
The advantage of such a model is that it
can be easily applied to nanotubes of different size, helicity,
and number of layers.
One such model has been used to predict the
phonon spectrum of small single-wall nanotubes.\cite{jishi2}
Here we present results of applying a similar model
to calculate elastic properties
of single and multi-wall nanotubes of various size
and geometry, and that of crystalline nanoropes composed of single-wall
nanotubes.

\vskip 0.2in
{\hspace{-0.5in}\large\bf II. The Force-constant Model}

In an empirical force-constant model, the
atomic interactions near the equilibrium structure
are approximated by
the sum of pair-wise harmonic potentials
between atoms.
In the most successful model for
the graphite, interactions
up to fourth-neighbor in-plane and out-of-plane
interactions are included.\cite{jishi}
The force constants are empirical determined
by fitting to measured elastic constants
and phonon frequencies.

The local structure of a nanotube layer
can be constructed from the conformal mapping of 
the graphitic sheet on to a cylindrical surface.
For a typical nanotube of few nm in radius,
the curvature is small enough that one expects
short-range atomic interactions to be 
very close to that in the
graphite. Thus, we adopt the same parameters developed
by Al-Jishi et al.\cite{jishi} for graphite
for intra-plane interactions in all nanotubes.

The different layers in a multi-wall nanotube 
are not well registered
as they are in the single crystal graphite.
Thus, one can not adopt the same set of parameters
for the interlayer interactions. Instead,
we model the interlayer interactions in nanotube
by the summation of pair-wise van de Waals interactions,
$U(r)=4\epsilon \left((\sigma/r)^{12}-(\sigma/r)^6)\right) $.
Such a model has been used successfully to calculate
the bulk properties of $C_{60}$ solid.\cite{lu92a}
The van de Waals parameter
$\sigma=3.4$\AA\ , $\epsilon=12$meV,
were determined by fitting the interlayer
distance and the elastic constant $c_{33}$ of the 
single crystal graphite.\cite{blakslee}

\vskip 0.2in
{\hspace{-0.5in}\large\bf III. Single-Layer Nanotubes}

Following the notation of White et al.\cite{white}, each single-layer
nanotube is indexed by a pair of integers $(n_1,n_2)$,
corresponding to a lattice
vector ${\bf L}=n_1 {\bf a_1}+n_2 {\bf a_2}$
on the graphite plane, where ${\bf a_1},{\bf a_2}$
are the graphite plane unit cell vectors.
The structure of the nanotube is obtained
by the conformal mapping of a graphite strip 
onto a cylindrical surface. 
The nanotube radius is given by
$R=a_o \sqrt{3(n_1^2+n_2^2+n_1 n_2)}/2\pi $,
where $a_o=1.42$\AA\ is the C-C bond length.

In principle, force constants depend
on the size of the nanotube as overlaps
of $\pi$ orbitals change with the nanotube
curvature.\cite{lu2}
However, Such dependence is very weak.
In this paper, we neglect this effect
and concentrate on the dependence of elastic properties
on the geometry and interlayer interactions.

The elastic constants are calculated from
the second derivatives of the energy density with
respect to various strains.\cite{huntington}
The tensile stiffness as measured by Young's modulus
is defined as the stress/strain ratio when a 
material is axially
strained. For most materials, the radial dimension
is reduced when it is axially elongated.
The ratio of the reduction in radial dimension
to the axial elongation
defines the Poisson ratio $\nu$.
We first calculate the Poisson ratio by minimizing the
strain energy with respect to both the radial compression and
the axial extension. The Young's modulus $Y$
is then calculated from the second derivative
of the strain energy density with respect to the axial
strain at the fixed $\nu$.

Table.1 lists the bulk, Young's and shear
(referred to the torsional shear) moduli calculated for 
selective examples of single-wall nanotubes.
An important quantity in determining values of
elastic constants is the wall thickness $h$ of
nanotubes.
Previous calculations has taken
$h=0.66$\AA\ for single layer nanotubes
which leads to the unusual large Young's modulus.\cite{bernholc}
For multi-wall nanotubes, all experiments indicate
that the interlayer distance is the same
as that in the graphite, $h=3.4$\AA . 
Thus, it is reasonable to take the interlayer distance
$h=3.4$\AA\ as the wall thickness.
We use the same values for all single multi-wall nanotubes.
This enables us to compare
results across nanotubes of different size and
number of layers.
For comparison, elastic moduli of
the graphite\cite{blakslee} and that of the diamond\cite{huntington}
are also listed.

Examine the numbers in Table.1 one concludes that:
(1) {\em Elastic moduli are insensitive to the
size and the helicity.} 
(2) {\em The Young's and shear moduli of nanotubes 
are comparable to that of the diamond and
that if in-plane graphite.}
(3) {\em Single-wall nanotubes are stiff in both the axial direction
and the basal plane.}

\vskip 0.2in
{\hspace{-0.5in}\large\bf VI. Multi-wall Nanotubes}

The interlayer distance in all 
experimentally observed multi-wall nanotubes
is comparable to that in graphite. 
This puts a constrain on possible combinations
of single-wall nanotubes to form
multi-wall nanotubes.
We have calculated elastic moduli for
many different combinations. It is found that
elastic properties are insensitive
to different combinations
as long as the constrain -- interlayer distance $\approx 3.4$\AA\ --
is satisfied.
Because of this insensitivity we use
results for one series of multi-wall nanotubes
to illustrate our main points.
The series chosen is constructed
from $(5n,5n), n=1,2,3\cdots$ 
single-wall tubes. This is one of the most likely
structure for multi-wall tubes as
its interlayer distance is very close
to that actually observed.\cite{ruoff}

Table.2 lists the calculated elastic coefficients
and the bulk, Young's, and shear modulus
for this series of nanotubes up to 10 layers.
The experimental values for the graphite and
the diamond are also listed for comparison.
One observes that
the elastic moduli are essentially
independent of the number of layers.
The same is true for all other multi-wall nanotubes
we have calculated. From Table.2 and its comparison
with Table.1 one concludes:
(1){\em The elastic moduli vary little with the number of layers.}
(2){\em The interlayer van de Waals interactions
contribute less than 10\% to the elastic moduli of
multi-wall nanotubes.}

The Young's modulus of multi-wall nanotubes was
deduced recently by Treacy et al.\cite{ebbesen}
from the thermal vibrations of anchored tubes.
Their values range from $0.4$ to $4$ TPa with
the average values of $1.6$ TPa.
These results are substantial larger
than our calculated values of $1$ TPa.
The discrepancy may be due to the large
uncertainty in how to estimate the Young's
modulus from their experiments.
In their estimation the isotropic model was assumed,
our results clearly show that this is not true.
More recent direct measurement of
the multi-wall nanotubes using the AFM technique
and the Euler's buckling criteria
has yield a result of $1$ TPa,\cite{dai} in agreement
with our calculations.

\vskip 0.2in
{\hspace{-0.5in}\large\bf V. Crystalline Nanoropes}

The discovery of a new efficient method of
producing bulk quantity of single-wall nanotubes
has made it possible to make
crystalline ropes of nanotubes.\cite{thess}
These nanoropes consist of 100 to 500 single-wall
nanotubes of uniform size. Due to the weak inter-tube
interactions one expects these rope to be flexible in the basal plane,
yet very stiff along the axial direction.

We use same model described above
to calculate the lattice constant and elastic moduli
of these nanoropes. 
Table.3 summarize the bulk properties of nanoropes
with nanotubes radius ranging from 1nm (the (5,5) tube)
to 2nm (the (13,13) tube).
Due to extreme disparity between
the inter-tube and intra-tube interactions,
we have neglected the coupling between the two interactions.
Thus, the lattice constant $a_0$ and the cohesive
energy $E_0$ are determined by inter-tube 
van de Waals interaction only.
It is found that $a_0$ and the
cohesive energy per atom scales with the tube radius $R$ as
$a_0=2R+3.2$\AA\ , $E_0=61.5 (meV)/\sqrt{R (\AA\ ) }$.
For nanorope composed of the
typical (10,10) tube, $R=6.78$\AA\ , $a_0=16.8$\AA\ , $E_0=23$meV.
The cohesive energy of the rope is comparable to
that of the $C_{60}$ solid (33mev).

From Table.3 one observes that:
(1) {\em Nanorope is very
anisotropic.}
(2) {\em The basal plane is soft, while the axial direction
is very stiff.}
(3) {\em The $C_{33}$ is about half that of the diamond.}
The weak inter-tube interaction ensures that 
the rope is flexible as individual tubes
can rotate and slide with respect to
each other easily.
This is supported by the experimental SEM images
where long nanoropes are observed
to be well bended and tangled.\cite{thess}

\vskip 0.2in
{\hspace{-0.5in}\large\bf VI. Conclusions}

We have investigated elastic
properties of nanotubes and nanoropes
using an empirical force constant model. The simplicity
of the model enable us to explore the dependence of
elastic moduli on the nanotube geometry.
It is shown that elastic properties
are insensitive to the radius,
helicity, and the number of layers.
The calculated Young's modulus ($\sim 1$ TPa)
and shear modulus ($\sim 0.5$ TPa)
are comparable to that of diamond for both
single and multi-wall individual nanotubes.
Crystalline rope of nanotubes is very
anisotropic in its elastic properties --
soft on basal plane and stiff along the axial
direction. The large Young's modulus and
flexibility of nanoropes make them ideal
materials for nanometer scale engineering.

\vskip 0.2in
{\em Acknowledgments} 
This work is supported by a grant from U.S.
Department of Energy, and in part by
a grant from The Petroleum Research Foundation.

\begin{table}
\vskip 1in
\vbox{
\caption{
Elastic moduli of selective single-wall nanotubes.
$(n_1,n_2)$ -- index, $R$ -- radius in nm.
$B,Y,M$ are bulk, Young's and shear modulus in
units of TPa ($10^{13} dy/cm^2$).
$\nu$ is the Poisson ratio.
Experimental values for the graphite and the diamond
are listed for comparison.
}
\vskip 0.5in
\begin{tabular}{||c|c|c|c|c|c||}
\makebox[0.8in]{ $(n_1,n_2)$}&\makebox[0.7in]{ $R$ }
&\makebox[1in]{ $B$}    &\makebox[1in]{ $Y$}  
&\makebox[1in]{ $M$}    &\makebox[1in]{ $\nu$ }\\
\hline
\hline
(5,5)    &  0.34  & 0.191  &  0.971  &  0.436  &  0.280  \\ \hline
(6,4)    &  0.34  & 0.191  &  0.968  &  0.437  &  0.284  \\ \hline
(7,3)    &  0.35  & 0.191  &  0.968  &  0.454  &  0.284  \\ \hline
(8,2)    &  0.36  & 0.190  &  0.974  &  0.452  &  0.280  \\ \hline
(9,1)    &  0.37  & 0.191  &  0.968  &  0.465  &  0.284  \\ \hline
(10,0)   &  0.39  & 0.192  &  0.968  &  0.451  &  0.282  \\ \hline
(10,10)  &  0.68  & 0.191  &  0.972  &  0.457  &  0.278  \\ \hline
(50,50)  &  3.39  & 0.192  &  0.969  &  0.458  &  0.282  \\ \hline
(100,100)&  6.78  & 0.192  &  0.969  &  0.462  &  0.282  \\ \hline
(200,200)& 13.56  & 0.192  &  0.969  &  0.478  &  0.282  \\ \hline
\hline
\multicolumn{2}{||c||}{ Graphite$^a$} &  0.0083
&  1.02   &  0.44 &  0.16 \\ \hline
\multicolumn{2}{||c||}{ Graphite$^b$} &  0.0083
&  0.0365 &  0.004 &  0.012 \\ \hline
\multicolumn{2}{||c||}{ Diamond$^c$} &  0.442
&  1.063  &  0.5758 &  0.1041 \\
\end{tabular}
\vskip 0.2in
$^a$ Graphite along the basal plane.\cite{blakslee} \\
$^b$ Graphite along the C axis.\cite{blakslee} \\
$^c$ Diamond along the cube axis.\cite{huntington}
}

\vbox{
\caption{Elastic coefficients and moduli (in TPa)
of multi-wall nanotubes
constructed from the $(5n,5n),n=1,2,3 \cdots $ series of 
single-wall tubes.
$N$ -- number of layers, $R$ -- radius of the out-most layer in nm.
$B, Y, M$ are bulk, Young's and shear modulus (in TPa).
Values for the graphite and the diamond are
listed for comparison.}
\vskip 0.5in
\begin{tabular}{||c|c|c|c|c|c|c|c|c|c||}
\makebox[0.3in]{\Large\bf {n}}       &\makebox[0.6in]{\Large\bf $R$}  &
\makebox[0.6in]{\Large\bf $C_{11}$}&
\makebox[0.6in]{\Large\bf $C_{33}$}&\makebox[0.6in]{\Large\bf $C_{44}$} &
\makebox[0.6in]{\Large\bf $C_{66}$} & \makebox[0.6in]{\Large\bf $C_{13}$} &
\makebox[0.6in]{\Large\bf $Y$}  &\makebox[0.6in]{\Large\bf $M$} &
\makebox[0.6in]{\Large\bf $B$} \\
\hline
\hline
  1 &  0.34  &  0.397  &  1.05  &  0.189  &  0.134  &  0.147  &  0.97 &  0.436  &  0.191 \\ \hline
  2 &  0.68  &  0.412  &  1.13  &  0.189  &  0.137  &  0.146  &  1.05 &  0.455  &  0.194 \\ \hline
  3 &  1.02  &  0.413  &  1.15  &  0.189  &  0.138  &  0.146  &  1.08 &  0.464  &  0.194 \\ \hline
  4 &  1.36  &  0.412  &  1.17  &  0.189  &  0.138  &  0.146  &  1.09 &  0.472  &  0.194 \\ \hline
  5 &  1.70  &  0.411  &  1.18  &  0.189  &  0.139  &  0.146  &  1.10 &  0.481  &  0.194 \\ \hline
  6 &  2.03  &  0.411  &  1.18  &  0.189  &  0.139  &  0.146  &  1.10 &  0.491  &  0.194 \\ \hline
  7 &  2.37  &  0.410  &  1.18  &  0.189  &  0.139  &  0.146  &  1.11 &  0.502  &  0.194 \\ \hline
  8 &  2.71  &  0.410  &  1.19  &  0.189  &  0.139  &  0.146  &  1.11 &  0.514  &  0.194 \\ \hline
  9 &  3.05  &  0.410  &  1.19  &  0.190  &  0.139  &  0.146  &  1.11 &  0.527  &  0.194 \\ \hline
 10 &  3.39  &  0.410  &  1.19  &  0.190  &  0.139  &  0.146  &  1.11 &  0.541  &  0.194 \\ 
\hline
\hline
\multicolumn{2}{||c|}{ Graphite$^a$}&1.06 &0.036 &0.004 &0.440 &0.015 &1.02& 0.008 &0.440 \\ \hline
\multicolumn{2}{||c|}{ Diamond} &1.07 &1.07  &0.575 &0.575 &0.125 &1.06& 0.442 &0.575
\end{tabular}
\vskip 0.2in
$^a$Young's and shear moduli refer to the basal plane. \\
}

\vbox{
\caption{Lattice constant $a_0$ (nm), cohesive energy per atom $E_0$ (meV),
and elastic moduli (in TPa)
of crystalline nanorope made of single-wall $(n,n)$ tubes.
$R$(nm) is the radius of single-wall tube.
}
\vskip 0.5in
\begin{tabular}{||c|c|c|c|c|c|c||}
\makebox[0.6in]{\Large\bf n}       &\makebox[0.9in]{\Large\bf $R$}  &
\makebox[0.9in]{\Large\bf $a_0$}     &\makebox[0.9in]{\Large\bf $E_0$}  &
\makebox[0.9in]{\Large\bf $C_{11}$}&\makebox[0.9in]{\Large\bf $C_{12}$} &
\makebox[0.9in]{\Large\bf $C_{33}$} \\
\hline
\hline
  5 &   0.33  &   0.99  &  33.5  &   0.066  &   0.022  &   0.795  \\ \hline
  6 &   0.40  &   1.13  &  30.1  &   0.071  &   0.024  &   0.736  \\ \hline
  7 &   0.47  &   1.26  &  28.2  &   0.078  &   0.024  &   0.687  \\ \hline
  8 &   0.54  &   1.40  &  26.2  &   0.082  &   0.029  &   0.641  \\ \hline
  9 &   0.61  &   1.54  &  24.7  &   0.085  &   0.029  &   0.600  \\ \hline
 10 &   0.67  &   1.67  &  23.5  &   0.090  &   0.032  &   0.563  \\ \hline
 11 &   0.74  &   1.81  &  22.5  &   0.098  &   0.035  &   0.532  \\ \hline
 12 &   0.81  &   1.94  &  21.6  &   0.102  &   0.036  &   0.502  \\ \hline
 13 &   0.88  &   2.08  &  20.7  &   0.106  &   0.036  &   0.475  \\ \hline
 14 &   0.94  &   2.21  &  19.9  &   0.111  &   0.042  &   0.452  \\ \hline
 15 &   1.01  &   2.35  &  19.3  &   0.118  &   0.043  &   0.430
\end{tabular}
}

\end{table}

\end{document}